\documentclass[useAMS,usenatbib]{mn2e}
\usepackage{graphicx}
\title[Period doubling and resonances in RR Lyrae models]{Period doubling bifurcation and high-order resonances in RR Lyrae hydrodynamical models}
\author[Z. Koll\'ath, L. Moln\'ar and R. Szab\'o]{Z. Koll\'ath\thanks{E-mail:
kollath@konkoly.hu (ZK); lmolnar@konkoly.hu (LM)}, L. Moln\'ar, R. Szab\'o \\
Konkoly Observatory, Budapest, 1121, Konkoly Thege \'ut 13-17, Hungary}
\begin{document}

\date{2011 January 26.}

\pagerange{\pageref{firstpage}--\pageref{lastpage}} \pubyear{2011}

\maketitle

\label{firstpage}

\begin{abstract}
We investigated period doubling, a well-known phenomenon in dynamical systems, for the first time in RR Lyrae models. These studies provide theoretical background for the recent discovery of period doubling in some Blazhko RR Lyrae stars with the \textit{Kepler} space telescope. Since period doubling was observed only in Blazhko-modulated stars so far, the phenomenon can help in the understanding of the modulation as well. Utilising the Florida-Budapest turbulent convective hydrodynamical code, we identified the phenomenon in radiative and convective models as well. A period-doubling cascade was also followed up to an eight-period solution confirming that the destabilisation of the limit cycle is indeed the underlying phenomenon. 

Floquet stability roots were calculated to investigate the possible causes and occurrences of the phenomenon. A two-dimensional diagnostic diagram was constructed to display the various resonances between the fundamental mode and the different overtones. Combining the two tools, we confirmed that the period-doubling instability is caused by a 9:2 resonance between the 9th overtone and the fundamental mode. Destabilisation of the limit cycle by a resonance of a high-order mode is possible because the overtone is a strange mode. The resonance is found to be sufficiently strong enough to shift the period of overtone with up to 10 percent. Our investigations suggest that a more complex interplay of radial (and presumably non-radial) modes could happen in RR Lyrae stars that might have connections with the Blazhko effect as well.
\end{abstract}


\section{Introduction}
Period doubling, a phenomenon often observed in dynamical systems, was also found in stellar models and actual pulsating variables in the last decades. Period doubling (PD) means that the observed quantity of the system alternates between a high and low amplitude cycle. Dynamical systems as the simple R\"ossler oscillator are usually capable of period doubling bifurcation, and through a series of bifurcations called the \textit{Feigenbaum cascade} can evolve to chaotic behaviour. The very definition of RV Tauri variables was originally the alternation of deep and shallow minima, a clear sign of PD (\textit{e.g.} \citealt{preston63}). The phenomenon was reproduced by \citet{fokin94} in radiative stellar models for RV Tauri stars as well as by \citet*{saitou89} in one-zone stellar models. Models of W Vir variables \citep{bk87} are also capable of bifurcation cascade towards chaos, and chaotic pulsations were indeed identified in semiregular stars (\citealt{bksm96}, \citealt{kbsm98}, \citealt*{bkc04}). Period doubling was reported by \citet{kiss02} in the Mira star R Cyg. Models of classical pulsators also showed promising results. \citet{mb90} searched for half-integer resonances both in Cepheid and RR Lyrae models and indeed found PD in the former case (see also \citealt{bm90}). A 2:3 resonance between the fundamental mode and the first overtone was also identified as a root cause. But neither period doubling nor suitable resonances were found in RR Lyrae stars between the fundamental or first overtone and any higher modes up to the fourth overtone. \citet{aikawa01} also reported PD in very long-period, radiative Cepheid models but did not identify any underlying resonance.

The signs of possible PD are the half-integer frequencies (HIFs) in the Fourier spectrum, in the form of $ (2n+1)/2\, f_0 $ with respect to the $f_0$ main periodicity, sometimes called as subharmonics. Similar peaks were found in some variable white dwarfs as well, suggesting PD in PG1351+489 \citep{goupil88} and even signs of four-period behaviour in G191-16 \citep{vauc89}. But because exact values usually differ slightly from half-integer values, they could be genuine non-radial modes that resonate with one of the higher amplitude modes (\textit{e.g.} \citealt{odonoghue92}).

\begin{figure*}
\includegraphics[width=180mm]{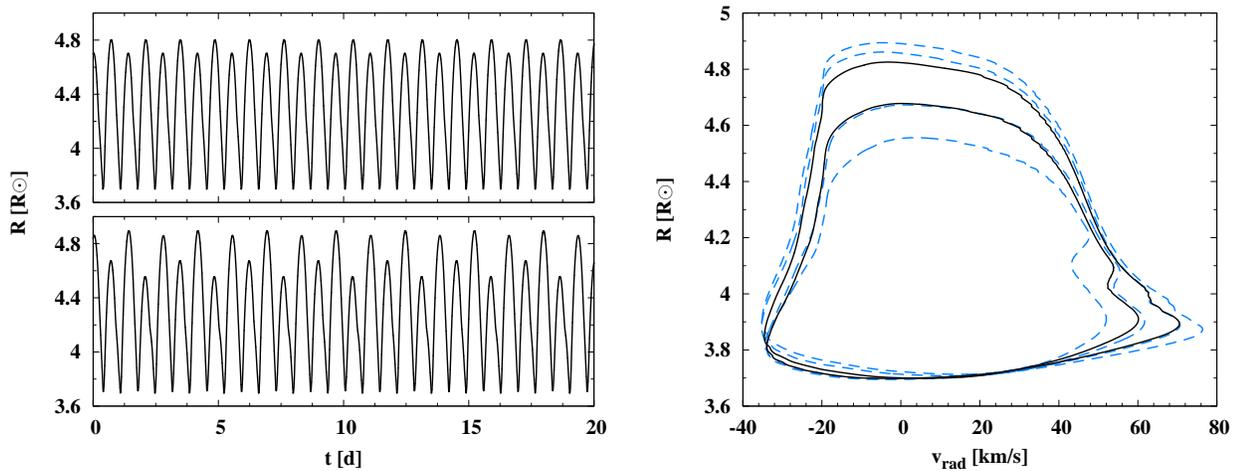}
\caption{Bifurcated solutions of a radiative model (6500 K, 0.57 M$_\odot$, 40 L$_\odot$). Period doubling is shown in the upper left, four-period variation on the lower left panel. The right-hand panel shows the same models in the radial-velocity--radius phase space, the black solid line is the double-period where the blue dashed line is the four-period solution. Both models were calculated with 100 mass shells and identical parameters except the value of the artificial viscosity. }
\label{p2p4} 
\end{figure*}

Period doubling in RR Lyrae variables was completely unexpected. As noted above, model calculations of \citet{mb90} did not yield positive results. No observational sign was ever reported either, beyond some residual scatter around pulsation maxima and minima (\citealt{kolenberg06}, \citealt{jurcsik08mwl}). With the typical pulsation periods spanning between 0.2-1 days, only multi-site campaigns have had real chance to discover period doubling, by covering consecutive pulsation cycles. But even an extensive observing run on RR Lyr did not resolve any additional frequencies beyond the pulsational, Blazhko and combination peaks or any other hints of period doubling (\citealt{kolenberg06}, 2011). More interestingly, PD had not been identified unambiguously in the CoRoT datasets yet, although \citet{poretti10} mention a weak peak at $1.5 f_0$ for the modulated RRab CoRoT 101128793. 

The first signs of alternating pulsation amplitudes were found in RR Lyr itself through the continuous and precise photometry of the \textit{Kepler} space telescope (\citealt{kolenberg10b}, \citealt{kepler}). \citet{pd} confirmed the case of RR Lyr and reported the discovery of period doubling in two other RR Lyrae stars while \citet{benko10} added four more \textit{Kepler} suspects to the list where weak HIF peaks are present in the frequency spectra. Interestingly enough, all seven stars where PD is present or presumed are Blazhko variables. Period doubling is dominantly present only at given Blazhko phases and even then the magnitude of the PD amplitude variations differ from one modulation cycle to another. All these facts considered, it is not that surprising that ground-based observations did not succeed. 

\citet{pd} gave detailed description of the \textit{Kepler} observations. It useful to recapitulate the basic observational facts regarding the period doubling in RR Lyrae stars: in the case of RR Lyr variation in subsequent maxima reaches $0.^m1$ but it amounts to a few hundredth of magnitude in the other stars. It is interesting to note that in some phases of the PD, especially for V808 Cyg, the minima are changing barely, however at other phases they show as large deviations from cycle to cycle as the maxima (see Fig. 6 in \citealt{pd}). The long cadence data hints the possibility of more complex features, \textit{e.g.} four-period structures but short cadence data will be needed to confirm it. \citet{pd} also reported that the magnitude of the PD variation and the amplitudes of most of the HIF peaks in the Fourier spectra change in concordance. The highest HIF peak is the $3/2 f_0$ in all three cases suggesting a 3:2 resonance, but such mechanism was excluded by \citet{mb90} for RR Lyrae stars. The amplitudes of consecutive peaks do not decrease smoothly: there is a bump around the $9/2 f_0$ peak, pointing towards the possibility of some higher order resonance. It was also suggested with hydrodynamical calculations that period doubling occurs because of a high-order resonance between the fundamental mode and the 9th overtone. In this paper we investigate the numerical calculations in detail to prove that the 9:2 resonance is indeed the underlying mechanism behind period doubling. 

\section{Period doubling in RR Lyrae models}
The period doubling phenomenon was identified in RR Lyrae model calculations by the Florida-Budapest turbulent convective hydrodynamical code. Throughout the calculations we used convective models with OPAL95 opacities ($Z=0.0001$ metal content) and 150 mass zones. The inner boundary temperature was set to 2.5 million K. Neither the stability of the static model nor the stability of the limit cycle solution depends significantly on the depth of the model. Increasing the inner boundary temperature up to 10 million K results only a few percent shift in the growth rates and stability roots. However, we preferred the lower inner boundary temperature as it provides better resolution in the models at the same number of zones. For details of the code we refer to \citet{kollath01} and \citet{kollath02}. The relaxation method, an effective way of reaching limit cycle solutions \citep{stellingwerf}, and the Floquet stability analysis \citep{floquet} are implemented and were used extensively in the calculations as well. 

Period doubling was first encountered in the models in the summer of 2009. Because of the complete lack of observational or theoretical evidence at the time, even with the space-based results of CoRoT, only low priority was given to the results. We recognised the importance of the PD models when the first batch of \textit{Kepler} data arrived \citep{kolenberg10b}. The initial hydrodynamical results were published along with the observations in \citet{pd}.

A bifurcated limit cycle is shown in the upper left panel of Figure \ref{p2p4}. Alternating maxima are pronounced, with $0.10\, R_\odot$ difference between cycles, while the effect is much smaller, only $0.002\, R_\odot$ at minima in this case. Period doubling occurs not just in convective but purely radiative hydrodynamical models as well. When calculating the radiative limit cases of convective models that show period doubling and using sufficient number of zones and low enough artificial viscosity (100 zones and $c_q<1.9$ in our case), the limit cycle becomes unstable and bifurcates to a double-period solution. If $c_q$ is decreased further, the model bifurcates to a four-period solution (lower left panel in Figure \ref{p2p4}; the period-doubling cascade will be discussed in Section \ref{bifcasc}). The prominent appearance of PD in radiative models raises the question why it was not reported before. In our own survey for double mode pulsation in \citet*{szaborrd}, we were interested only in the transient behaviour of the amplitude evolution of the modes and did not investigate the limit cycles. 

Because period doubling in Cepheids was connected to the half-integer resonances, we set out to investigate them in the RR Lyrae models as well. A large grid of linear models were created to identify the occurrences of various resonances. As lower-order overtones were excluded by \citet{mb90}, modes up to the 14th overtone were all calculated. But handling of the large number of possible resonances with the fundamental mode required additional tools.

\subsection{Diagnostic diagram}
Models with a given $P_0/P_k$ resonance spread out to a 2D surface in the three-dimensional $T_{eff}$--$M$--$L$ (effective temperature--mass--luminosity) space. Analysing multiple near-resonance regions is not straightforward in such cases. We found however that these surfaces are not twisted in all directions: a useful two-dimensional projection of the 3D space can be constructed where the resonant models are confined to a narrow region. With this projection one looks ``edge-on'' at the surfaces. The relation we used is $T_{eff}$ \textit{vs.} $150 M - L$ where the mass and luminosity values are in solar units. The exact value of the mass coefficient is possibly different for each resonances but since none of them reduces to a 1D line, we used the value 150 for convenience. Our calculations cover a more extended range of masses of luminosities compared to the physical range of stars, in order to follow all resonances throughout the $150M -L$ parameter.

\begin{figure}
\includegraphics[width=85mm]{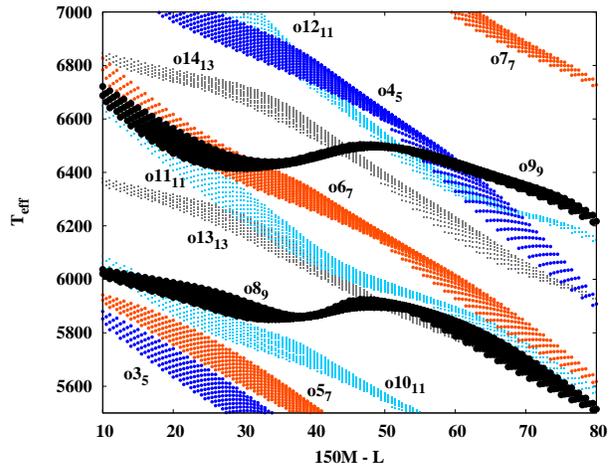}
\caption{The diagnostic diagram. Resonance regions are plotted in a projection of the original $T_{eff} - M - L$ space, the x axis being $150M-L$. Overtone and resonance orders are labelled in $\mathrm{o}k_l$ form where $k$ is the overtone order and the $l$ lower index is the $P_0/P_k = l/2$ value. The strange modes $o8_9$ and $o9_9$ are quite distinguishable from the normal modes. Different colours represent different resonance orders.}
\label{diagn} 
\end{figure}

The regions for different eigenmodes and resonance values are plotted in Figure \ref{diagn}. Occurrences of resonances with normal modes are monotonically changing whereas resonance regions of strange modes are curved. This difference comes from the same behaviour that is noted at the modal diagram too (see Section \ref{strange}): normal modes follow an ordered and well separated series with fairly similar period ratios that explains the similar slopes while strange modes deviate and have changing slopes.

There are a number of other resonances with different overtones close to the 9:2 resonance of the 9th overtone, two of which are lower order than the strange mode. We concentrate on excluding these two as the higher-order ($k > 10$) normal modes are even less likely to have any effect on the fundamental one. The 7:2 resonance region of the 6th overtone and the 5:2 resonance region of the 4th overtone are well separated and distinctly different from the strange mode. Therefore model sequences covering a wide enough range of $150 M - L$ values could discriminate between the modes and select the one that causes the bifurcation. To achieve this, the stability properties of the models had to be investigated.

\subsection{Strange modes}
\label{strange}
As mentioned earlier, no suitable resonances were found between the pulsational mode and some overtone up to the fourth order in previous studies. Higher order radial overtones were usually disregarded because they normally are heavily damped and thus have no measurable effect on the pulsation. In special cases though high overtone modes deviate from this simple picture, in the form of strange modes.

Strange modes were first identified in highly non-adiabatic systems such as luminous helium stars where \citet{wood76} found large differences between adiabatic and non-adiabatic mode sequences. \citet{cox80} also reported additional non-adiabatic modes in models with high luminosity--mass ratios ($L_\odot / M_\odot \sim 10^4$).  The term ``strange mode'' was introduced by \citet{cox80} simply to name the additional mode among the lower pulsational modes. It was originally thought that strange modes require strong nonadiabaticity, but they were also identified later in weakly non-adiabatic Cepheid and RR Lyrae models \citep{bk01}. Strange modes were eventually found in classical pulsators \citep*{byk97} and adiabatic models of massive stars \citep*{kiri93} as well. Both studies concluded that strange modes are essentially acoustic waves that are trapped by a barrier in the form of a large sound speed gradient inside the star. Partial ionisation zones can act as very efficient boundaries that effectively decouple the star into two separate regions, both having their own oscillation spectra. Strange modes are therefore modes that are trapped between the barrier and the stellar surface and usually have peculiarly high growth-rates. Although strange modes can be identified in adiabatic models as well, the differences are much more emphasized in the non-adiabatic models. Despite the extensive theoretical work however, there is no unique and precise definition yet. For more details on strange modes we refer to \citet*{saio98}.

\begin{figure}
\includegraphics[width=85mm]{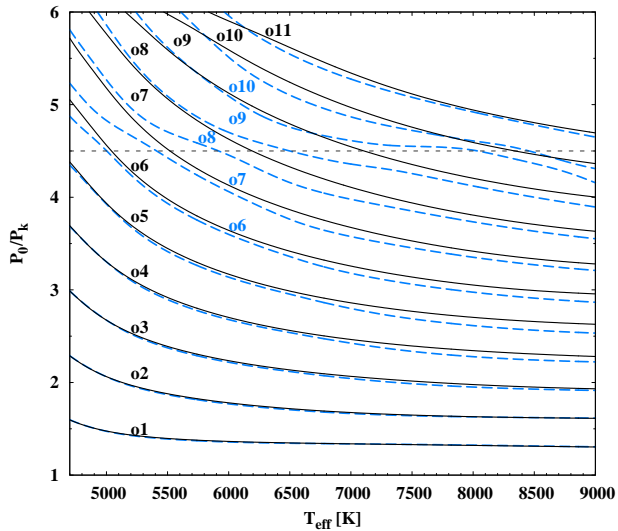}
\caption{Modal diagram of an RR Lyrae model sequence. Periods are scaled with the fundamental mode. Black solid lines denote the adiabatic modes whereas blue dashed lines denote the non-adiabatic modes. The 9/2 period ratio is indicated with the horizontal dashed line. Deviations are most pronounced around this period ratio, indicating the presence of strange modes.}
\label{strangefig} 
\end{figure}

\begin{figure}
\includegraphics[width=80mm]{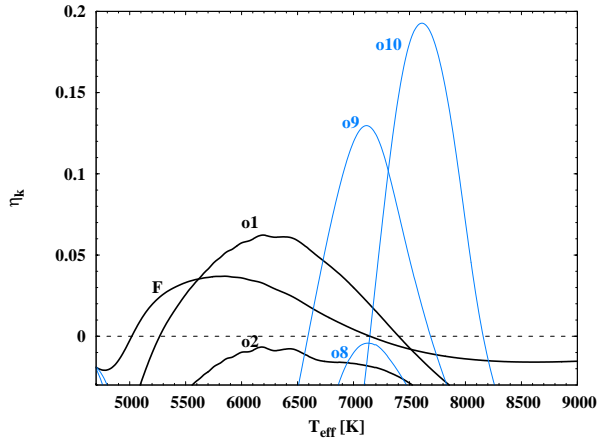}
\caption{Linear growth rates of radial modes. Starting with the second overtone, modes become more and more damped, but then overtone eight gets close to the unstable regime, indicating its strangeness. Overtone nine and ten could become excited, as well.}
\label{gr} 
\end{figure}

Strange modes can be identified on the modal diagram of our RR Lyrae model sequence in Figure \ref{strangefig}. Adiabatic (black solid lines) and non-adiabatic (blue dashed lines) align well up to the sixth overtone. But around $9/2 f_0$ (or $4.5 \, P_0/P_k$ on the figure) modes appear without evident adiabatic pairs, switching from overtone seven to ten at higher effective temperatures through avoided crossings. The ninth and tenth overtones can have large linear growth rates too (Figure \ref{gr}) but positive values are largely above the blue edge of the RRab instability region and mostly outside the identified PD region. The overtone region 8-10 was found to be close to the instability limit by \citet{glasner93} too. After the identification of strange modes in classical pulsators, it was shown by \citet{byk97} that the behaviour of growth rates can be explained by the occurrence of strange modes. We observe the same excursion towards instability too, as illustrated in Figure \ref{gr}. But we are less interested in linear growth rates than the effects of nonlinear interaction between the strange mode and the fundamental mode.

\section{Stability of the limit cycle}
The stability of the limit cycle and the presence of a destabilising strange mode is described by the Floquet analysis. For the required calculations we refer to \citet{stellingwerf} and \citet*{bmk91}. We give only a brief summary of the method here: after reaching a limit cycle solution, we perturb each independent variable in the model. Then the evolution is followed for one period and the difference from the unperturbed limit cycle is approximated through a Jacobi matrix. The stability of the limit cycle is then given by the eigenvalues of the matrix in the form of $F_k = \mathrm{exp} (\lambda_k + i \phi_k)$ where $F_k$ are the $k$th order Floquet coefficients and $\lambda_k$ and $\phi_k$ are the corresponding exponents and phases. For the stability analysis of the limit cycle, we used our turbulent convective code. Then the variables in the Floquet matrix are the radius, velocity, turbulent energy  and total energy of the zones. However, the results obtained by radiative models (where turbulent energy is ignored) do not differ significantly from the turbulent ones, as the additional eigenvalues related to turbulent energy do not play a role in the destabilization of the fundamental mode limit cycle. In general, the addition of turbulent energy in the linear stability calculations introduces only damped thermal modes. Low-order vibrational modes can be associated with the corresponding Floquet coefficients as $\phi_k \approx P_0 / P_k (\mathrm{mod}\: 2\pi)$ and  $\lambda_k \approx \kappa_k P_0$, if the system is close to the blue edge of the instability strip. But mode identification becomes more and more complicated, especially for higher-order modes when the model gets farther from the blue edge, because $\phi_k$ itself shifts and close to resonances the phase locks to $\pi$ (or zero) for any overtone. 

\citet{mb90} showed that in the vicinity of resonances the usually complex Floquet coefficients bifurcate into a real pair, often creating a bubble in the corresponding exponents when plotted against $T_{eff}$: the two real values first diverge then converge to the complex meeting point as the control parameter is varied. The corresponding phases discriminate between period ratios: for integer resonances ($nP_k \approx P_0$, n small and integer) the Floquet phase is $\phi_k=0$ while for half-integer resonances ($(2n+1)/2 P_k \approx P_0$) the phase is $\phi_k = \pi$ during the split of exponents. Period doubling occurs in the latter case when one of the exponents (the upper arm) becomes positive. 

Because of the more rapid variations of higher order coefficients a dense grid was required in all control parameters. Model sequences with selected $150 M - L$ values and with 25-50 $K$ steps in effective temperatures were calculated. We only considered the eddy viscosity ($\alpha_{\nu}$) from the convective parameters as it does not change the equilibrium stellar structure but can have measurable effects on the stability of the limit cycle. At each temperature step $\alpha_{\nu}$ was varied from 0.04 to 0.1 (or even from $\alpha_{\nu}\approx 0$ to 0.1 in some cases). Such wide range in $\alpha_{\nu}$ is not expected in real stars but was required to identify the coefficients correctly. The step size was set to $\Delta\alpha_{\nu} = 2 \cdot 10^{-4}$, which translates into 3--500 individual models for each and every mass, luminosity and effective temperature value. Because the equilibrium stellar structure is not modified when $\alpha_{\nu}$ is changed slightly, the model can be relaxed from the previous limit cycle solution to the new one directly. Still, a large amount of CPU time was required to map the PD instability region. Our calculations took several months on a 4-processor Intel Core2 Quad 2.83 GHz machine but a similar analysis using the turbulent convection strength parameter $\alpha_c$ would require far more as presumably every single model would have to be iterated close to the limit cycle first and then relaxed in every step instead.

\subsection{Period doubling and resonances}
\begin{figure}
\includegraphics[width=85mm]{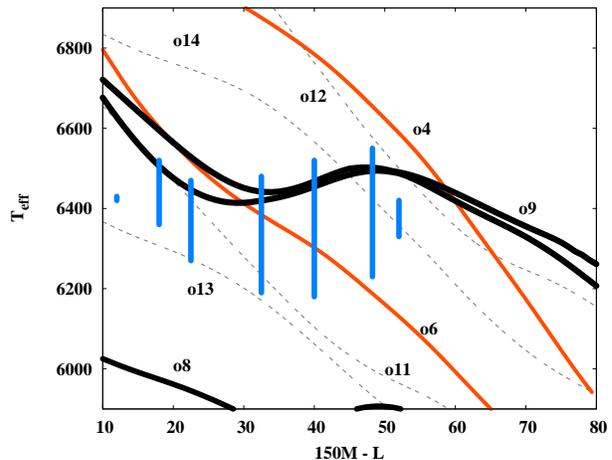}
\caption{Models where the resonant, positive Floquet exponent is observed at $\alpha_{\nu}=0.05$, \textit{e.g.} period doubling is present plotted over the diagnostic diagram (vertical blue columns). The models follow the 9:2 resonance of the 9th overtone. Only mean values of the resonance bands are plotted (grey dashed lines), except for the 9th overtone where the edges are shown (thick black lines). The two lower-order normal modes and the the 8th overtone strange mode are plotted with orange and black lines respectively. }
\label{pd_map} 
\end{figure}

Figure \ref{pd_map} shows a simple map of the PD instability region plotted on the diagnostic diagram. The selection was based on the Floquet spectra: the plotted models exhibit a positive resonant exponent at $\alpha_\nu = 0.05$. At low $150 M-L$ values the resonance regions of overtone six and nine overlap or are close to each other and the same is true for the fourth overtone at higher values. The sixth overtone crosses the PD region but no positive exponents are observed at lower or higher temperatures and the PD region is almost completely separated from the fourth overtone. On the other hand, the non-monotone strip of the 9:2 resonance of the ninth overtone clearly follows the PD region. It is interesting to note that the centre of the PD region is shifted to lower effective temperatures than the place of the actual linear resonance of the ninth overtone. We will discuss this shift in more detail in section \ref{shift}. Still, we consider the 9:2 resonance the best explanation since the other resonances are much more separated from the margins of the PD region.

\subsection{Bifurcation cascade}
\label{bifcasc}
\begin{figure}
\includegraphics[width=85mm]{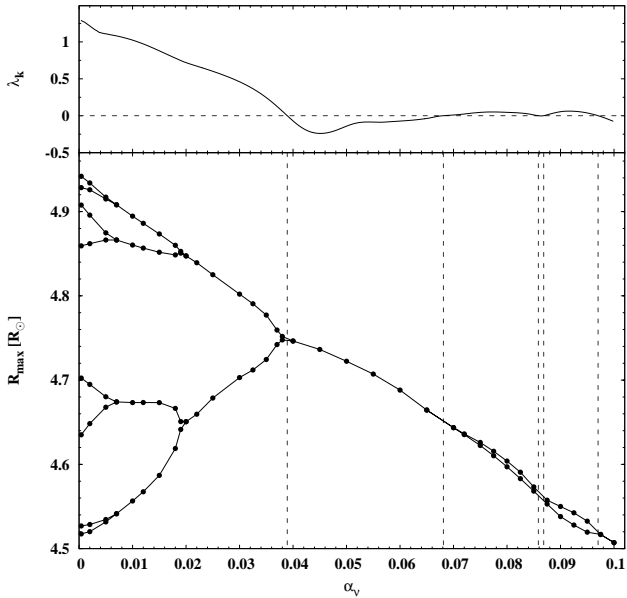}
\caption{Bifurcation cascade of an RR Lyrae model ($T_{eff} = 6500 K, M = 0.59 M_\odot, L=56 L_\odot$). Each model correspond to a limit cycle solution with different $\alpha_\nu$ eddy viscosity parameters. The plotted values are the maximum stellar radii of all different cycles. The upper panel shows the corresponding Floquet exponent. When the exponent is positive, the limit cycle is bifurcated. The vertical lines indicate the successive zero crossings of the exponent. }
\label{bifurc} 
\end{figure}

Successive period doubling events create a bifurcation cascade that could lead even to chaos. Our models show a period doubling cascade as well: Figure \ref{bifurc} displays the multiplication of the number of different pulsation maxima versus the $\alpha_\nu$ eddy viscosity parameter. Towards the lower values the model bifurcates to an eight-period solution through double- and four-period stages. The cascading we observed in both radiative and turbulent convective models confirms that a bifurcation cascade can arise in the limit cycle itself. Alternating amplitudes, resembling period doubling, might be generated through specific conditions as well \citep{fokin94}. This phenomenon however is a true bifurcation in a dynamical sense, which is confirmed by the cascade. We note again however that models with very low eddy viscosity parameters are interesting from the dynamical point of view: they do not represent actual stars.

It is interesting to note in Figure \ref{bifurc} that period doubling occurs at two different $\alpha_\nu$ ranges. The cascade is prominent on the left-hand side, at $\alpha_\nu < 0.04$. Then the model bifurcates again around $\alpha_\nu \sim 0.07$. The behaviour of the limit cycle is in perfect agreement with the value of the corresponding Floquet exponent that drops from large positive values to negative with increasing $\alpha_\nu$ but grows back to small positive between $0.069 < \alpha_\nu < 0.097$ thus generating small-amplitude period doubling there.

\subsection{Examples of Floquet spectra}
The calculated Floquet spectra revealed surprisingly complex behaviour. Rapid variations in the exponents corresponding to higher order ($k > 5$) modes required a fine step-rate in the $\alpha_{\nu}$ parameter as discussed above. Then the exponents were sorted: values from one edge were followed along the sequence to identify individual $\lambda_k (\alpha_{\nu}))$ datasets. Two different selection criteria were set: any sets containing $\phi_k = 0$ resonant phases were filtered out while sets containing $\phi_k \simeq \pi$ were differentiated from the remaining non-resonant exponents. With this method, no resonant exponents were missed. If one calculates the models for a single $\alpha_{\nu}$ value only, exponents related to resonances could be missed because the $\phi_k \simeq \pi$ condition is valid only for a limited range of $\alpha_{\nu}$ in some cases.

\begin{figure}
\includegraphics[width=85mm]{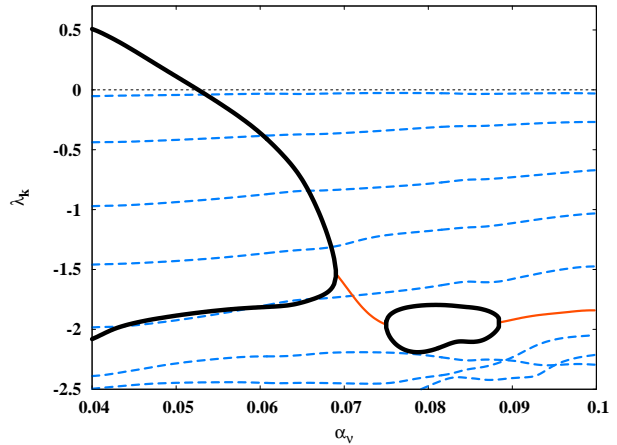}
\caption{Floquet exponents versus the eddy viscosity parameter. Resonant values ($\phi_k = \pi$) are shown with thick black lines, the connecting resonant, not phase-locked intervals ($\phi_k < \pi$) are shown with orange solid lines. Dashed blue lines represent other exponents with $0 < \phi_k < \pi$. Exponents where $\phi_k = 0$ appeared were excluded for clarity. One coefficient becomes strongly unstable for lower eddy viscosities ($\alpha_\nu<0.055$), creating PD in the limit cycle. Exponents at low values start to cross each other. Model parameters are: $T_{eff} = 6400 K,\, M=0.6 M_\odot,\, L=38L_\odot$.}
\label{floq1} 
\end{figure}

A relatively simple example is shown in Figure \ref{floq1}. The first top few dashed lines could be identified with the lowest-order modes. At lower values however, the exponents start to cross each other (they are separated in phase of course), making the further identification almost impossible: the ninth overtone for example is shifted to higher values than expected because of the resonance. The exponent we are interested in is plotted with black and orange lines: black means the phase-locked (real) solutions with $\phi_k=\pi$ (split into two arms) that are connected with intervals with complex Floquet exponents (orange lines). The value of the exponent varies rapidly, confirming that a single model with a badly chosen $\alpha_{\nu}$ would indeed miss the unstable ($\alpha_\nu<0.055$) or even the real valued solutions.

Not all Floquet spectra are that easy to explain: we encountered not just splitting into two arms but more complex structures suggesting some connection and/or interaction between individual resonances. This is especially true for models with higher effective temperatures where most of the exponents vary more rapidly. However, these findings do not affect our conclusions on period doubling and will be the scope of future research.

\begin{figure}
\includegraphics[width=85mm]{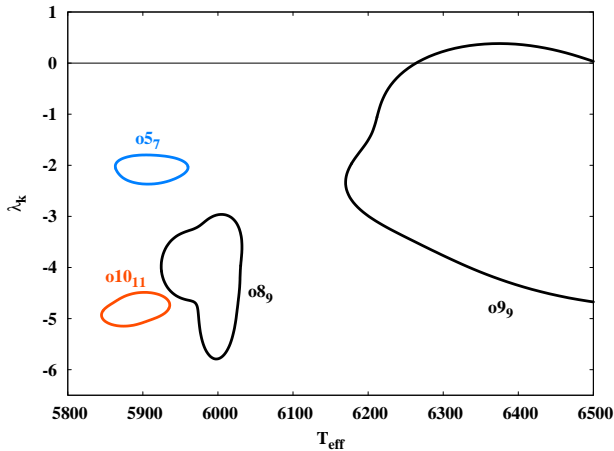}
\caption{Schematic diagram of the resonant Floquet exponents versus effective temperature. Model parameters are: $150 M-L=22.5$, $\alpha_{\nu}\sim 0.06$. The most likely overtones and resonances are indicated. The limit cycle is destabilised if the exponents turn positive.}
\label{floq_vst} 
\end{figure}

We attempted to reconstruct the exponents that contain resonant, phase-locked portions in the $\lambda_k$ vs. $T_{eff}$ plane for a given $\alpha_\nu$ value, to identify various bubbles with the corresponding resonances. This turned out to be less straightforward as we expected because of the various interactions between the exponents, making the identification along the $T_{eff}$ values either difficult or ambiguous. We also note that the exact shapes of the bubbles depend strongly on the chosen $\alpha_\nu$ value as seen in Figure \ref{floq1}. Instead we created a schematic diagram (Figure \ref{floq_vst}), showing the positions and approximate sizes of the bubbles, labelled with the most likely resonances. The diagram displays the strong presence of the two strange modes (o$8_9$ and o$9_9$) and the negative exponents of the resonances except of the upper arm of o$9_9$, further confirming that the o$9_9$ resonance is the only possible explanation behind the period doubling. The high-order resonances with negative Floquet exponents do not have any significant effect on the  pulsation, most probably they cannot be observed. The highest order resonance known which results in observable distortion of the light-curve with a negative Floquet exponent is the 1:2 resonance of the fundamental mode and the 4th overtone in s-Cepheids \citep*{feucht00}. 

\section{Discussion}
\subsection{Shift from the linear resonance}
\label{shift}
The identified PD region clearly shows some shift regarding the resonance region of the strange mode, as stated above. One has to bear in mind though that period ratios were calculated from linear eigenvalues: actual non-linear periods will differ from those values. The difference can be easily computed for the fundamental pulsational mode from the period of the limit cycle. Calculation of non-linear high-overtone periods cannot be done so easily, except for the PD region: because these are resonant models, the phase lock of the Floquet coefficients to $\pi$ guarantee that the period of the ninth overtone is exactly $P_0/4.5$. In our set of models, the period of the fundamental mode is increased by $\approx 0.002$ d, approximately the same value for all parameter values. The non-linear period shift of the ninth overtone has the same order in the centre of the PD region but it has a higher variation with temperature.
At the edge of the PD instability, the strange mode can shift by up to $\sim 0.01$ d or $10 \%$ in respect of the linear value ($P_9 \sim 0.1$ d) because of the resonant coupling, indicating the surprising strength of this particular 9:2 resonance. 

We have to note however that the properties of the strange mode or surface mode depends on the boundary conditions and zoning. Numerical uncertainties can also play a role in the shift of periods. The difference between equilibrium and limit cycle models can also change the period of the strange mode.

\subsection{Effects of metallicity}
\begin{figure}
\includegraphics[width=85mm]{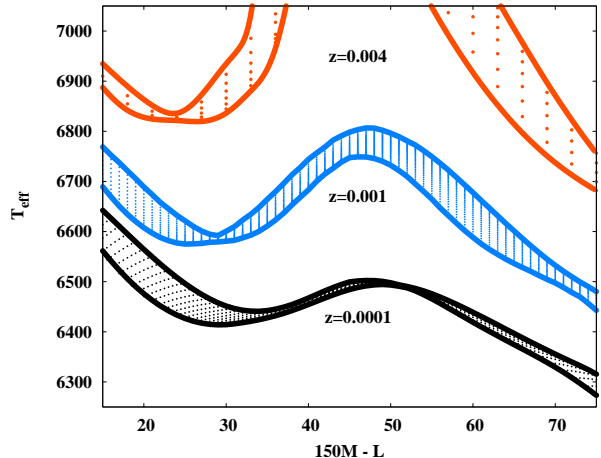}
\caption{Location of the 9:2 resonance of the ninth overtone on the diagnostic diagram for different metallicities. Black: $Z=0.0001$, blue: $Z=0.001$, orange: $Z=0.004$. Points are individual model calculations, curves represent the upper and lower envelopes of the $T_{eff}\,(150M-L)$ projection of the surface of resonant models. Note that for $Z=0.004$ the intermediate $150M-L$ values are devoid of the particular resonance.}
\label{diagz} 
\end{figure}

Metal content in the models was set to $Z=10^{-4}$ but field RR Lyrae stars are sometimes more metal-abundant. We calculated linear models with $Z=0.001$ and $0.004$ values too. The appearance of the 9:2 resonance in the diagnostic diagram changed in two ways (see Figure \ref{diagz}): first, it shifted to higher effective temperatures, secondly, it became more extended and curved. The gap in the $Z=0.004$ band is not a plotting artefact but a real phenomenon. There are no resonant models for the intermediate $150M-L$ values (between about 40 and 50), even up $T_{eff} \le 7200 K$, the period ratio of the fundamental mode and the strange mode avoids the 9:2 value.

One can readily assume that the PD instability region shifts to higher temperatures as well. If the PD region is about as narrow for higher metallicities as it is for $Z=0.0001$, observing PD in stars with independently measured effective temperature and metallicity values could provide a simple test of this mechanism. 

\subsection{The case of RR Lyr}
Out of the three \textit{Kepler} Blazhko-modulated stars showing the period doubling effect, only RR Lyr has effective temperature, luminosity and mass values published already. An effective temperature of $T_{eff} = 6480 \pm 120 K$ was derived from multicolour photometry by \citet{siegel82}. JHK near-infrared time-series photometry yielded the following parameters: $T_{eff} = 6330 \pm 50 K$, $L = 48 \pm 5 L_\odot$ and $M = 0.63 \pm 0.01 M_\odot$, using a $Z = 0.0012$ metallicity value \citep{sollima}. By means of high-resolution spectroscopy, \citet{kolenberg10a} derived metallicity between $Z=0.001 - 0.003$ and also calculated a luminosity of $L=49 \pm 5 L_\odot$. The mass and luminosity values translate to a $150M-L* \cong 46 \pm 6$ value. 

RR Lyr would fit perfectly into our PD instability region (Figure \ref{pd_map}) if it had lower metallicity by an order of magnitude. Although we did not calculate nonlinear models with higher metallicities, RR Lyr could also fit in the $Z=0.001$ PD instability. Linear resonances occur between $6700-6800 K$ in this case. The calculated PD region has a width of $\sim 300 K$ below the linear resonance: extrapolating this relation, RR Lyr would fall to the edge $(\sim 6400 K)$ of the PD instability of higher metallicity. These estimates suggest the the higher effective temperature measurements $(T_{eff} > 6400K)$ are preferred for RR Lyr from the PD point of view. We cannot however exclude numerical uncertainties as the cause of discrepancy between the position of RR Lyr and the PD region.

\section{Conclusions and future work}
We presented the analysis of the period doubling phenomenon we encountered in RR Lyrae hydrodynamical models. The behaviour of the models match the observed period doubling in modulated \textit{Kepler} RR Lyrae stars \citep{pd}. Our investigations revealed the following:
\begin{enumerate} 
 \item The underlying phenomenon is the destabilisation of the fundamental mode limit cycle. Not only period doubling was observed: a cascade up to eight-period solutions was also followed.
 \item The root cause behind the period doubling is a 9:2 resonance between the fundamental mode and the ninth overtone. This is the first example of such a high-order resonance that is able to destabilise the limit cycle. The ninth overtone itself was found to be a strange mode. The resonance is strong enough to shift the period of the overtone as large as 10 percent.
 \item Floquet analysis was found to be a crucial (and also time-consuming) method to connect period doubling to the 9:2 resonance, together with our diagnostic diagram ($150M - L$ vs. $T_{eff}$). 
 \item The strength of period doubling is sensitive to the convective parameters of the models. In this investigation we considered only the eddy viscosity parameter, there are however others, most notably the strength of the turbulent convective flux and/or the mixing length.
 \item Increasing the metallicity shifts the resonance (and thus the PD instability) to considerably higher effective temperatures. If confirmed, this relation could help narrowing down the metal content of other RR Lyrae stars showing period doubling.
\end{enumerate}

There are about 30 known RR Lyrae stars in the \textit{Kepler} field, half of which do not exhibit Blazhko modulation. No PD was found however in any of the non-modulated stars even below mmag level \citep{pd}. Yet the models have no problems producing period doubling in non-modulated stars. Different factors could explain this hiatus. For example the Kepler sample might be too small; HB evolutionary models would shed more light on the population of the PD instability region and on the time span the stars spend in and out of it; the modulated stars cover a wider parameter space that make the conditions required for PD easier to occur. 

The results suggest that PD can indeed play an important role in the understanding of the Blazhko effect. But is period doubling a cause or an effect? If the modulation is indeed caused by some kind of internal variation of the stellar structure, like in the Stothers model \citep{stothers06}, then PD might be another manifestation of this mechanism. Assuming that physical parameters vary over the Blazhko cycle (see \citealt{sodor09} and references therein), these stars could sweep over a wider parameter range, providing better circumstances for the PD instability to occur. 

Since PD is a resonance phenomenon, resonances could also be connected to the origin of the Blazhko modulation (see \citealt{kovacs09} for a recent summary on the models). Thus purely geometrical explanations as the magnetic oblique rotator model \citep{shibahashi00} are not well suited since there is no direct connection between the modulating process, the rotation, and the modal resonances. On the other hand, the non-radial resonance model \citep{nd01} links the modulation to a 1:1 resonance between the fundamental and a non-radial mode. As the PD instability demonstrates it, high-order modes can have a strong effect on the pulsation, suggesting that other resonances or some more complex interplay of radial and non-radial modes and resonances might occur in the stars. On the other hand, the same resonance that is responsible for the period doubling can result in different kinds of bifurcations. The amplitude equation formalism provides possibilities to extend the hydrodynamical calculations to a more general study. These theoretical investigations may provide new theories for Blazhko phenomenon as stated in a companion paper \citep{bk11}. All these new developments point towards the possibility of a complicated mechanism behind the Blazhko effect that may include resonances or variations in the stellar structure or even the two processes simultaneously.

The fact that high-order resonances with strange modes play an important role in the observed behaviour of stellar pulsation arises lots of new questions. We have started extended hydrodynamical surveys to test the effect of these resonances for other types of radial pulsators and for different parameter sets. The resonance of only two modes provides period doubling, three-mode resonances however (\textit{e.g.} a resonance among two low-order modes and a high-order strange mode) can also provide interesting effects like three-mode pulsation. Predictions for such modal interactions will be discussed in a forthcoming paper.

\section*{Acknowledgments}
Fruitful discussions with J. R. Buchler are gratefully acknowledged. We thank the referee for his/her helpful comments that helped to improve the paper. This work has been supported by the Hungarian OTKA grants K83790 and MB08C 81013, as well as the `Lend\"ulet' program of the Hungarian Academy of Sciences.

\bsp

\label{lastpage}

\end{document}